\documentclass{JHEP3}
\usepackage[dvips]{epsfig}
\usepackage{graphicx,amssymb,amsmath,amsfonts,cite}
\usepackage{bm}
\usepackage{verbatim}

\usepackage[english]{babel}

\usepackage{bm}

\title{Lifshitz Hydrodynamics}

\author{Carlos Hoyos, Bom Soo Kim, Yaron Oz\\
Raymond and Beverly Sackler School of
Physics and Astronomy, Tel-Aviv University, Tel-Aviv 69978, Israel\\
E-mail: \email{choyos,bskim,yaronoz@post.tau.ac.il}}

\abstract{
We construct the hydrodynamics of quantum critical points with Lifshitz scaling. There are new dissipative effects allowed by the lack of boost invariance. The formulation is applicable, in general, to any fluid with an explicit breaking of boost symmetry. We use a Drude model of a strange metal to study the physical effects of the new transport coefficient. It can be measured using electric fields with non-zero gradients, or via the heat production when an external force is turned on. Scaling arguments fix the resistivity to be linear in the temperature. 
}

\preprint{TAUP-2964/13}

\begin{document}

\section{Introduction and summary}

Heavy fermion compounds and other materials including high $T_c$ superconductors have a metallic phase (dubbed as `strange metal') whose properties cannot be explained within the ordinary Landau-Fermi liquid theory. In this phase some quantities exhibit universal behaviour such as the resistivity, which is linear in the temperature $\bm{\rho} \sim T$ \cite{PhysRevLett.59.1337,PhysRevLett.85.626,2013Sci...339..804B}. 
Such simple and universal scaling properties of transport coefficients and thermodynamic quantities are believed to be the consequence of quantum criticality \cite{Coleman:2005,Sachdev:2011,2008NatPh...4..186G}, 
where scaling symmetries impose strong constraints on its dynamics even without well-defined quasiparticle or tractable microscopic descriptions. 

At the quantum critical point, there is a Lifshitz scaling  \cite{Hornreich:1975,Grinstein:1981} symmetry that affects differently to time and space directions
\begin{equation}  \label{ScalingTR}
t \rightarrow \Omega^z t,~~~~~x^i \rightarrow \Omega x^i,~~~ i=1,...d  \ .
\end{equation}
For special values of the `dynamical exponent' $z\!=\! 1 $ and $ 2$, the spacetime symmetry can be enhanced to include the Lorentz and the Galilean groups, respectively. In both cases the extra symmetries include transformations between inertial frames (i.e.~moving at relative constant velocity) or boosts. 
Lorentz boosts are the transformations 
\begin{equation}
( t,x^i)\to \frac{1}{\sqrt{1-v^2/c^2}}\left( t-\frac{v_i}{c^2} x^i,\,x^i-v^i t\right),
\end{equation}  
where $c$ is a maximal velocity ($v^2\leq c^2$). Galilean boosts can be obtained from 
\begin{equation}
(t,x^i)\to (t, \, x^i-v^i t).
\end{equation}
For all other values of $z$, boost symmetries will be generically broken in either case. One can distinguish both cases because in the `Galilean' case there is a conserved mass density and in the `Lorentzian' case a maximal velocity.

It is well known that systems with ordinary critical points behave hydrodynamically with transport coefficients whose temperature dependence is determined by the scaling at the critical point \cite{Hohenberg:1977}. Quantum critical systems also have hydrodynamic descriptions, as has been shown more recently for conformal field theories at finite temperature \cite{Kovtun:2004de}, fermions at unitarity \cite{Cao:2011} and graphene \cite{Mueller:2008,Fritz:2008,Mueller:2009}. A similar hydrodynamic description has been suggested for strange metals based on the large scattering rate measured in experiments \cite{PhysRevLett.69.2411,Zaanen,2013Sci...339..804B}. The hydrodynamic expansion is very universal but it can be constrained by symmetries and other physical requirements, leading to distinct predictions for different classes of theories. 

Despite its obvious interest for the description of strange metals, the corresponding hydrodynamic description for quantum critical points with Lifshitz scaling has not been formulated yet. In contrast to the previous examples such a description should take into account the effects due to the lack of boost invariance.  As first argued in \cite{PhysRevLett.69.2411}, the hydrodynamic description of quantum critical points will be appropriate if the characteristic length of thermal fluctuations $\ell_T \sim 1/T^{1/z}$ is much smaller than the correlation length $\xi\gg\ell_T$, which is the case for a large region of the phase diagram. If the size of the system $L$ is smaller than the correlation length then deviations from criticality will be unimportant, but in the hydrodynamic approximation we should also demand that gradients are much smaller than the temperature $\xi\gg L \gg\ell_T$. 
%

	\begin{figure}[!ht]
	\begin{center}
	\includegraphics[width=4in]{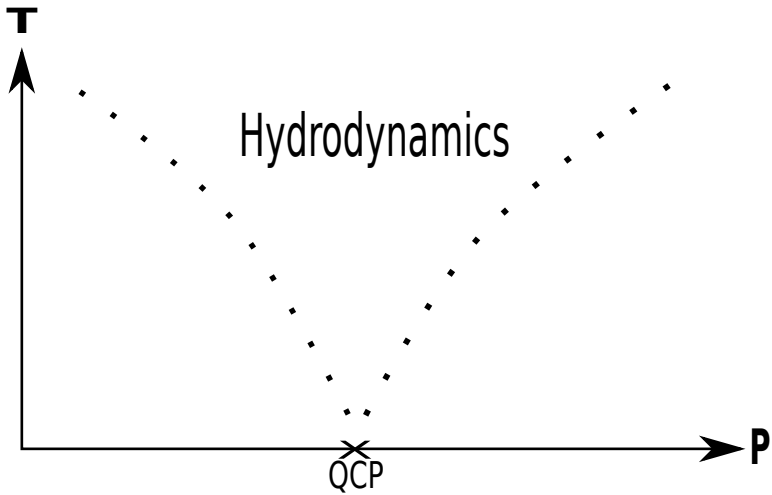}  
	\caption{Valid region of hydrodynamic description of quantum critical point (QCP). 
	$T$ is a temperature and $P$ is a quantum tuning parameter. }
	\end{center}
	\end{figure} 

In this paper we construct the hydrodynamic description of quantum critical points with Lifshitz scaling to first viscous order. Our results are universal up to the value of the coefficients in the hydrodynamic expansion, which depend on the details of the critical point. The hydrodynamic expansion depends on whether the broken boost symmetry belongs to the Lorentz or the Galilean group. We study both cases. 

Our main new result is the discovery of a single new transport coefficient allowed by the absence of boost invariance. Due to it, when the fluid is moving non-inertially there are new dissipative effects. The result applies to any system with Lifshitz scaling, but also more generally to any system where boost invariance is explicitly broken. For instance, fluids moving through a porous medium or electrons in a dirty metal.

For the application to strange metals, we assume that the Galilean description is the appropriate one, and use a hydrodynamic model as an effective description to the long wavelength collective motion of the electrons. We compute the conductivity using a Drude model similar to the one used for graphene in \cite{Mendoza:2013}.
The effect of the new coefficient is manifested as a non-linear dependence on an applied electric field. Interestingly, we also find that scaling arguments fix the resistivity to be linear in  temperature, under a reasonable assumption that it is linear in the mass density. This behaviour is universal: it is independent of the number of dimensions and the value of the dynamical exponent. The result is not strictly new because a linear behaviour was also derived in the context of marginal Fermi liquids \cite{PhysRevLett.63.1996}, but it is worth mentioning because previous hydrodynamic derivations using a Lorentzian description led to a different temperature dependence \cite{Hartnoll:2009ns}.
 
\section{Lifshitz Hydrodynamics}
We start considering the `Lorentzian' case and will take the non-relativistic limit later to obtain the `Galilean' fluid. The scaling dimension of the temperature is fixed by the dynamical exponent $[T]=z$. Simple dimensional analysis fix the scaling dimensions of the energy density and the pressure to be $[\varepsilon]=[p]=z+d$. Since the temperature is the only scale, the scaling completely fixes their temperature dependence
\begin{equation}
\varepsilon \sim p \sim T^{\frac{z+d}{z}}\,.
\end{equation}
The equation of state of a fluid with Lifshitz symmetry 
\begin{equation}
z\varepsilon = dp\,,
\end{equation}
then follows from the first law of thermodynamics $\varepsilon+p=Ts$, where $s=\partial p/\partial T$ is the entropy density. 

\subsection{Lifshitz algebra} 

One can also formally derive the equation of state from a Ward identity associated to the generator of Lifshitz scale transformations. The generators of Lifshitz symmetry are time translation $P_0=\partial_t$, spatial translations $P_i=\partial_i$, the scaling transformation 
$D=-z t\partial_t - x^i \partial_i$ and rotations. The subalgebra involving $D$, $P_i$ and $P_0$ has commutation relations 
\begin{align}
&[D, P_i] = P_i  \;, \quad [D,P_0]= z P_0 \;. 
\end{align}
In a field theory the scaling symmetry is manifested as a Ward identity involving the components of the energy-momentum tensor
\begin{equation}\label{wardid}
z T^0_{\ 0}+\delta^j_{\ i}T^i_{\ j}=0 \ .
\end{equation}
At finite temperature $T^0_{\ 0}=-\varepsilon$ is (minus) the energy density and $T^i_{\ j}=p\delta^i_{\ j}$ is the pressure, leading to the equation of state
\begin{equation}\label{eos}
z\varepsilon=d p \ .
\end{equation}
This fixes the temperature dependence of energy and pressure. Taking the dimension of spatial momentum to be one, the scaling dimensions are
\begin{equation}
[T]=z \ , \quad  [\varepsilon]=[p]=z+d\ .
\end{equation}

The Lifshitz algebra can be generalized for constant velocities $u^\mu$, $u^\mu u_\mu=\eta_{\mu\nu}u^\mu u^\nu=-1$ ($\mu,\nu=0,1,\cdots,d$), with scaling dimension $[u^\mu]=0$. We define the generators
\begin{equation}
P^\parallel= u^\mu \partial_\mu, \ \ P^\perp_\mu=P_\mu^{\ \nu}\partial_\nu, \ \ D=z x^\mu u_\mu P^\parallel-x^\mu P^\perp_\mu .
\end{equation}
Where $P^{\ \nu}_{\mu}=\delta^{\ \nu}_{\mu}+u_\mu u^\nu$. Then, the momentum operators commute among themselves and
\begin{equation}
[D,P^\parallel]=z P^\parallel \ , \quad [D,P^\perp_\mu]=P^\perp_\mu \ .
\end{equation}
The Ward identity associated to $D$ becomes
\begin{equation}\label{wardu}
z T^\mu_{\ \nu} u_\mu u^\nu-T^\mu_{\ \nu} P_\mu^{\ \nu}=0 \ .
\end{equation}
It coincides with \eqref{wardid} only when $z=1$, but leads to the equation of state \eqref{eos} for any velocity. The equation of state is thus independent of interactions. One obtains the same expression in strongly coupled theories like the holographic models proposed in  \cite{Kachru:2008yh,Koroteev:2007yp} as a gravitational dual to Lifshitz points, and we will detail the calculation elsewhere \cite{ComingPaper}. 

\subsection{Hydrodynamics}

The conservation of the energy-momentum tensor determines the hydrodynamic equations 
\begin{equation}
\partial_\mu T^{\mu\nu}=0.
\end{equation}
Lorentz symmetry forces the energy-momentum tensor to be symmetric. If boost or rotational symmetries are broken this condition can be relaxed. This allows many new terms in the hydrodynamic energy-momentum tensor, but as usual there are ambiguities in the definition of the hydrodynamic variables in the constitutive relations. In order to fix them, we impose the Landau frame condition
\begin{equation}\label{landau}
T^{\mu\nu}u_\nu=-\varepsilon u^\mu \ .
\end{equation}
Then, the generalized form of the energy-momentum tensor is
\begin{align}
T^{\mu\nu} = &(\varepsilon+p) u^\mu u^\nu+p \eta^{\mu\nu} +\pi_S^{(\mu\nu)}+\pi_A^{[\mu\nu]} +(u^\mu\pi_A^{[\nu\sigma]}+u^\nu\pi_A^{[\mu\sigma]})u_\sigma \ . \label{tmn}
\end{align}
We use the notation where round brackets denote symmetrization and square brackets antisymmetrization. The first line is the ideal part of the energy-momentum tensor. $\pi_S$ and $\pi_A$ denote all other possible terms that can appear in a derivative expansion of the velocity and the temperature. The condition \eqref{landau} implies the constraint $\pi_S^{(\mu\nu)} u_\nu=0$.  To first dissipative order
\begin{equation}
\pi_S^{(\mu\nu)}=-\eta^{\mu\nu\alpha\beta}\partial_\alpha u_\beta=-\eta P^{\mu\alpha} P^{\nu\beta} \Delta_{\alpha\beta}-\frac{\zeta}{d} P^{\mu\nu}\partial_\alpha u^\alpha,
\end{equation}
where $\eta$ and $\zeta$ are the shear and bulk viscosities respectively. The shear tensor is defined as
\begin{equation}
\Delta_{\alpha\beta}=2\partial_{(\alpha} u_{\beta)}-\frac{2}{d}P_{\alpha\beta}(\partial_\sigma u^\sigma).
\end{equation}
The antisymmetric term $\pi_A$ has no additional constraints for the last term in \eqref{tmn} ensures that the condition \eqref{landau} is satisfied. In a theory with rotational invariance $\pi_A^{[ij]}=0$. 

The new terms should be compatible with the laws of thermodynamics, in particular with the second law. In its local form it implies that the divergence of the entropy current must be semi-positive definite
\begin{equation}\label{secondlaw}
\partial_\mu j_s^\mu\geq 0.
\end{equation} 
The divergence of the entropy current can be derived from the conservation equation $\partial_\mu T^{\mu\nu}u_\nu=0$, which is
\begin{align}
T\partial_\mu(s u^\mu)=-\pi_A^{[\mu\sigma]}(\partial_{[\mu} u_{\sigma]}-u_{[\mu} u^\alpha \partial_\alpha u_{\sigma]})+\cdots.
\end{align}
Here $s$ is the entropy density and $T$ the temperature. Both are related to the energy density and the pressure through the first law of thermodynamics $\varepsilon+p=T s$.
In the Landau frame we can define the entropy current as $j_s^\mu=s u^\mu$ to first dissipative order. The dots denote positive-definite contributions proportional to the shear and bulk viscosities. 

In order for the antisymmetric contribution to be positive, we should be able to write it as a sum of squares. 
The positivity condition can be satisfied only if 
\begin{equation}
\pi_A^{[\mu\nu]}=-\alpha^{\mu\nu\sigma\rho}(\partial_{[\sigma} u_{\rho]}-u_{[\sigma} u^\alpha \partial_\alpha u_{\rho]}) \ ,
\end{equation}
where $\alpha^{\mu\nu\sigma\rho}$ contains all possible transport coefficients to first dissipative order. It is analogous to the viscosity tensor $\eta^{\mu\nu\sigma\rho}$, but instead of being symmetric on the first and last pair of indices it is antisymmetric. It must also satisfy the condition, for an arbitrary real tensor $\tau_{\mu\nu}$,
\begin{equation}
\tau_{\mu\nu}\alpha^{\mu\nu\sigma\rho}\tau_{\sigma\rho} \geq 0 \ .
\end{equation}
The coefficients $\alpha^{\mu\nu\sigma\rho}u_\mu P_\nu^{\ \alpha} u_\sigma P_{\rho}^{\ \beta}\neq 0$ break boost invariance and $\alpha^{\mu\nu\sigma\rho}P_\mu^{\gamma} P_\nu^{\ \alpha} P_\sigma^{\ \delta} P_{\rho}^{\ \beta}\neq 0$ rotational invariance in the fluid rest frame.
 
If only boost invariance is broken, there is a single possible transport coefficient $\alpha^{\mu\nu\sigma\rho}=\alpha u^{[\mu}P^{\nu][\rho}u^{\sigma]}$, with $\alpha \geq 0$:
\begin{equation}
\pi_A^{[\mu\nu]}=-\alpha u^{[\mu} u^\alpha \partial_\alpha u^{\nu]} \ .
\end{equation}
For a theory with Lifshitz symmetry the scaling dimension of the transport coefficients is $[\eta]=[\zeta]=[\alpha]=d$, which  determines their temperature dependence to be
\begin{equation}
\eta\sim\zeta\sim \alpha \sim T^{\frac{d}{z}} \ .
\end{equation}

There may be additional transport coefficients in a theory with more conserved charges. 
Indeed, we find two more transport coefficients to first dissipative order in the case with a conserved global current \cite{ComingPaper}. We also derive the Kubo formulas for the new transport coefficients in \cite{ComingPaper}.

\subsection{Non-relativistic limit}

We now study fluids with broken Galilean boost invariance. In the relativistic fluid the maximal velocity $c$ appears in $u^\mu=(1,\beta^i)/\sqrt{1-\beta^2}$, where $\beta^i=v^i/c$. In the non-relativistic limit $c\to\infty$, the pressure is not affected while the relativistic energy is expanded in terms of the mass density $\rho$ and the internal energy $U$ as
\begin{equation}
\varepsilon=c^2\rho-\frac{\rho v^2}{2}+U \ .
\end{equation}
The relativistic hydrodynamic equations reduce to the non-relativistic form 
\begin{align}
&\partial_t\rho +\partial_i(\rho v^i)=0 \ ,\\
&\partial_t U+\partial_i\left(U v^i\right)+p\partial_i v^i =\frac{\eta}{2}\sigma^{ij}\sigma_{ij}+\frac{\zeta}{d}(\partial_i v^i)^2+\alpha(V_A^i)^2 \ ,\\
 \notag &\partial_t(\rho v^i)+\partial_j(\rho v^j v^i)+\partial^i p \\ 
&\qquad\quad  =\partial_j\left(\eta\sigma^{ij}+\frac{\zeta}{d}\delta^{ij}\partial_k v^k\right)
  +\partial_t(\alpha V_A^i)+\partial_j\left(\alpha v^jV_A^{i}\right) \ .
\end{align}
The shear tensor is $\sigma_{ij}=\partial_i v_j+\partial_j v_i-(2/d)\delta_{ij}\partial_k v^k$. While taking the limit, we have absorbed factors of $1/c$ in the shear and bulk viscosities $\eta$ and $\zeta$ and a factor $1/c^3$ in $\alpha$. The vector $V_A^i$ is 
\begin{equation}
V_A^i=D_t v^i=(\partial_t+v^k\partial_k)v^i\ ,
\end{equation}
the relative acceleration of the fluid. Similarly to the viscosities, the coefficient $\alpha$ determines the dissipation that is produced in the fluid when the motion is not inertial.

In the non-relativistic limit with a non-zero mass density $\rho\neq 0$ the scaling symmetry needs to be modified. Under a space-time diffeomorphism 
\begin{equation}
t\to t+\xi^t \ , \ \ x^i\to x^i+\xi^i \ ,
\end{equation}
the partition function of the theory will change as
\begin{equation}
\delta \log Z = \int dt d^d x\left(-\partial_\mu\xi^t j^\mu_\varepsilon+\partial_\mu \xi^i T^\mu_{\ i}  \right) \ .
\end{equation}
Where $j_\varepsilon^\mu$ is the energy current, $T^t_{\ i}$ the momentum density and $T^i_{\ j}$ the stress tensor. If the transformation associated to $\xi^\mu$ is a symmetry, the variation of the partition function should vanish, leading to a Ward identity.

The Lifshitz equation of state is recovered if the theory has a symmetry
\begin{equation}
\xi^t=zt \ , \quad \xi^i=x^i+\frac{z-2}{2}v^i t \ .
\end{equation}
This is a combination of a scaling transformation \eqref{ScalingTR} and a change of frame. When $z=2$ the transformation is independent of the velocity and the symmetry group can be extended to include Galilean boosts and non-relativistic conformal transformations. The Ward identity becomes
\begin{equation}
0=-z j^t_\varepsilon+ \sum_i T^i_{\ i}+\frac{z-2}{2}v^i T^t_{\ i}=-z U+dp \ .
\end{equation}

In a fluid with Lifshitz symmetry the scaling dimensions of the hydrodynamic variables are
\begin{equation}
[v^i]=z-1, \quad  [p]=[U]=z+d, \quad [\rho]=d+2-z,
\end{equation}
while the temperature has scaling dimension $[T]=z$. We can determine the scaling dimensions of the transport coefficients by imposing that all the terms in the hydrodynamic equations have the same scaling. We find 
\begin{equation}
 \ \ [\eta]=[\zeta]= d \ , \quad [\alpha]=d-2(z-1) \ .
\end{equation}

\section{Drude model of strange metals}
We model the collective motion of electrons in the strange metal as a
charged fluid moving through a static medium, that produces a drag on the fluid.
%

The hydrodynamic equations are 
\begin{align}
\partial_\mu J^\mu=0, \quad
\partial_\mu T^{\mu 0}=J^i E_i, \quad
\partial_\mu T^{\mu i}=J^0 E^i-\lambda c J^i \ .
\end{align}
Note that we consider the case where the magnetic field is zero and $\partial_0 E_i=0$. 
We are interested in describing a steady state where the fluid has been accelerated by the electric field, increasing the current until the drag force is large enough to compensate for it. We assume that the flow does not change, but some scalar quantities like the energy can change with time. In order to simplify the calculation we consider only an incompressible fluid $\partial_i v^i=0$, which is valid when the velocities are much smaller than the speed of sound. The fluid motion is described by the Navier-Stokes equations
\begin{align}\label{navstok}
&\rho v^k\partial_k v^i+\partial^i p
 =\rho E^i-\lambda \rho v^i+\eta\nabla^2 v^i+\alpha \partial_j\left(v^jv^k\partial_k v^i\right) \ .
\end{align}
We have added two new terms: the force produced by the electric field $E^i$, 
and a drag term, whose coefficient $\lambda$ has scaling dimension $[\lambda]=z$. Both are expected to be present in the description of electrons moving through the medium  \cite{Mendoza:2013}. If the drag term was absent, momentum would be conserved in the absence of external forces and one expects on general grounds an infinite DC conductivity or, more precisely, a delta function contribution to the AC conductivity. Non-conservation of momentum, or equivalently, breaking of translation invariance is thus necessary to have a finite DC conductivity. See \cite{2007PhRvB..76n4502H,Hartnoll:2012rj} for other works 
where this topic is discussed in more detail.

\subsection{Conductivity} 

We can solve this equation order by order in derivatives. For constant pressure $\partial^ip=0$ and an external electric field%
\footnote{ If the pressure is not constant it will simply add up to the electric field. More generally, one could add other forces that will combine with the electric field in the same fashion. In order to make the presentation simpler we keep the pressure constant.},
we find the current satisfies Ohm's law to leading order 
\begin{equation}
J^i=\rho v^i \simeq \frac{\rho}{\lambda} E^i \ ,
\end{equation}
and the conductivity is simply $\bm{\sigma}_{ij}=\rho/\lambda\delta_{ij}$. The coefficient $\lambda$ has units of inverse time and it is proportional to the resistivity. Comparison with experimental values determines it to be linear in the temperature and of the order of the inverse `Planckian' dissipation time $\lambda \!\sim\! k_B T/\hbar$ \cite{PhysRevLett.69.2411,Zaanen,2009Sci...323..603C,2013Sci...339..804B}. We will see below how the linear dependence on the temperature follows from Lifshitz scaling.
%

Note that the form of the term proportional to $\alpha$ in \eqref{navstok} implies that the contribution to the conductivity will depend on gradients of the electric field squared. This is independent of all the simplifications we have made. For large enough gradients of the electric field the effect will be visible. Whether this is experimentally realizable depends on the magnitude of $\alpha/\rho$.

At higher orders in derivatives we find the following corrections for a divergenceless electric field $\vec{E}=E_y\hat{y}+E_x(y)\hat{x}$, where $E_y$ is constant,
\begin{equation}
\bm{\sigma}_{xx}(E_x, E_y) = \frac{\rho}{\lambda}\left[1+\frac{1}{\rho \lambda }\left(\eta +\left[\frac{\alpha }{\lambda} + \frac{\rho}{\lambda^3} \right] E_y^2\right)\frac{\partial_y^2 E_x}{E_x}-\frac{1}{\lambda^2}\frac{E_y\partial_y E_x}{E_x}\right] \ .
\end{equation}
The conductivity depends on the electric field and its gradients. 
When the electric field takes the form $E_x=E_0 \cos(y/L)$, the contribution of $\alpha$ to the conductivity is $y$ dependent
\begin{equation}
\bm{\sigma}_{xx}(E_x, E_y) = \frac{\rho}{\lambda}\left[1-\frac{1}{\rho \lambda }\left(\eta +\left[\frac{\alpha }{\lambda} + \frac{\rho}{\lambda^3} \right] E_y^2 \right)\frac{1}{L^2}-\frac{1}{\lambda^2}\frac{E_y\partial_y E_x}{E_x}\right] \ .
\end{equation}
If we average on the $y$ direction, we find that the conductivity decreases with the magnitude of the transverse electric field
\begin{equation}
\bm{\sigma}_{xx}(E_x, E_y) = \frac{\rho}{\lambda}\left[1-\frac{1}{\rho \lambda }\left(\eta +\left[\frac{\alpha }{\lambda} + \frac{\rho}{\lambda^3} \right] E_y^2  \right)\frac{1}{L^2}\right] \ .
\end{equation}
For completeness, we estimate the term proportional to $\alpha$ in the conductivity by comparing with the result for $\partial_y E_x=0$
\begin{align}
\frac{\delta\bm{\sigma}_{xx}-\bm{\sigma}_{xx}^0}{\bm{\sigma}_{xx}^0}\sim  10^{-11}\left(\frac{m_e}{m_*}\right)^2 \left(\frac{T}{{\rm K}}\right)^{-3}\frac{\alpha/\rho}{{\rm sec}}(\partial E_0)^2.
\end{align}
Where $m_*$ is the mass of the charge carriers, $m_e$ the electron mass and $ (\partial E_0) = \frac{E_0/L}{{\rm N}\,{\rm m}^{-1}\, {\rm C}^{-1}} $.

\subsection{Linear resistivity in temperature}

In contrast with a relativistic fluid, the density is approximately independent of the temperature. This introduces an additional scale, and in general the transport coefficients can be non-trivial functions of the ratio $\tau=T^{\frac{d+2-z}{z}}/\rho$. The conductivity will have the following temperature dependence 
\begin{equation}
\bm{\sigma}_{xx}=  T^{\frac{d-2(z-1)}{z}}\hat{\bm{\sigma}}(\tau)\simeq \frac{\rho}{T} \ ,
\end{equation}
where we assumed a linear dependence on the density as obtained from the calculation with the drag term. Note that this predicts a resistivity linear in the temperature and {\em independent} of the dynamical exponent $z$ and the number of dimensions $d$. 

It would be interesting to extract other quantities from the model. An observable that is also measured in experiments is the Hall angle, whose temperature dependence is $\cot \theta_H \sim T^2$ \cite{PhysRevLett.67.2088}. This behaviour is not as straightforward to obtain. The na\"ive scaling from this model would be $\cot \theta_H \sim T$. The scaling could be different if there is a strong temperature dependence of the permeability, while the permittivity is approximately constant.

\subsection{Dissipative effects}
We now study the heat production due to the introduction of external forces and the drag. Energy dissipation due to a drag force is also considered in the context of holographic models of Lifshitz theories in \cite{Kiritsis:2012ta}.

An electric field or temperature gradient will induce an acceleration
\begin{equation}
a^i=-\partial^i p/\rho+E^i=(s/\rho)\partial^i T+E^i \ .
\end{equation}
We impose $\partial_t a^i=0$, $\partial_j a^i=0$. The Navier-Stokes equations for homogeneous configurations takes the form
\begin{equation}
\partial_t v^i-(\alpha/\rho)\partial_t^2 v^i+\lambda v^i=a^i \ .
\end{equation}
If the forces are suddenly switched on at $t=0$, the evolution of the velocity is determined by this equation with the initial conditions $v^i(t=0)=0$, 
$\partial_t v^i(t=0)=\frac{ \rho a^i}{2 \alpha  \lambda }  \left(\sqrt{\frac{4 \alpha  \lambda }{\rho
   }+1}-1\right).$
\begin{equation}
\partial_t v^i(t=0)=\frac{ \rho a^i}{2 \alpha  \lambda }  \left(\sqrt{\frac{4 \alpha  \lambda }{\rho
   }+1}-1\right) \ .
\end{equation}
This choice is based on the physical requirement that at large times the velocity stays constant. When $\alpha\to 0$ it simply becomes $\partial_t v^i(t=0)=a^i$. 

The heat production rate induced by the force is
\begin{equation}
\partial_t U= \lambda\rho v^2+\alpha(\partial_t v)^2 \ .
\end{equation}
At late times the system evolves to a steady state configuration with constant velocity, so the heat production rate becomes constant $v^i=a^i/\lambda$. Subtracting this contribution for all times, the total heat produced is
\begin{equation}
\Delta Q =
 -\frac{\rho a^2}{2 \lambda ^2}  \left(\sqrt{4  \lambda \alpha  /\rho
   +1}+2\right) \ .
\end{equation}
The coefficient multiplying $\alpha/\rho$ is of order $4\lambda \sim 10^{12}\,{\rm sec}^{-1}\, (T/{\rm K})$. The  overall coefficient in the heat per unit mass $\Delta Q/\rho$ is of order $\sim (1\, {\rm m}/{\rm sec})^{2}\,(T/{\rm K})^{-2} (m_e/m_*)^2 (E_i/({\rm N}\, {\rm C}^{-1}))^2$.

Since the units and the scaling dimension of $\alpha/\rho$ are the same as $1/\lambda$, a possible guess is that it is proportional to the Planckian dissipation time $\alpha/\rho \sim 10^{-11}\, {\rm sec}\, (T/{\rm K})^{-1}$. Although this is quite a small number, it can reduce $\Delta Q$ to a fraction of its value compared to when $\alpha=0$. However, we do not have a justification for this choice from a microscopic point of view, it is possible that $\alpha/\rho$ contains terms that are independent of the temperature.

\section{Acknowledgements}
We would like to thank J.~de Boer, C.~Herzog, K.~Jensen, E.~Kiritsis, R.~Loganayagam, M.~M\"uller, C.~Panagopoulos, G.~ Policastro, M.~Rangamani and A.~Yarom for discussions. This work is supported in part by the Israeli Science Foundation center
of excellence.

\bibliographystyle{JHEP.bst}   
\bibliography{LifshitzBib.bib}

\end{document}